# Tungsten silicide films for microwave kinetic inductance detectors

Thomas Cecil[1], Antonino Miceli[1], Orlando Quaranta[1], Chian Liu[1], Daniel Rosenmann[2], Sean McHugh[3], and Benjamin Mazin[3]

1) X-ray Science Division, Argonne National Laboratory, Argonne, IL 60439, USA
2) Center for Nanoscale Materials, Argonne National Laboratory, Argonne, IL 60439, USA
3) Department of Physics, University of California, Santa Barbara, CA 93106, USA

Microwave Kinetic Inductance Detectors provide highly multiplexed arrays of detectors that can be configured to operate from the sub-millimeter to the x-ray regime. We have examined two tungsten silicide alloys ($W_5Si_3$ and $WSi_2$), which are dense alloys that provide a critical temperature tunable with composition, large kinetic inductance fraction, and high normal-state resistivity.  We have fabricated superconducting resonators and provide measurement data on critical temperature, surface resistance, quality factor, noise, and quasiparticles lifetime. Tungsten silicide appears to be promising for microwave kinetic inductance detectors.

Microwave Kinetic Inductance Detectors (MKIDs)[1] have gained widespread use as photon detectors for radiation covering the electromagnetic spectrum from the sub-millimeter[2] to X-rays[3]. In its simplest form an MKID consists of a thin film superconductor patterned into a microwave resonator that offers near unity transmission off resonance and a dip in transmission on resonance. MKIDs are generally operated at temperatures well below the superconducting transition in order to minimize the thermal population of quasiparticles. When a photon is absorbed in the superconductor it breaks Cooper pairs creating an excess of non-equilibrium quasiparticles. This excess of quasiparticles alters the complex surface impedance of the superconductor, raising the kinetic inductance and surface resistance, which causes the resonance feature to shift to a lower frequency and broaden. When the resonator is driven at a constant frequency, the change in kinetic inductance can be seen as a motion tangential (phase) and normal (amplitude) to the resonance circle. Since the phase signal is often much larger than the amplitude signal (despite having more noise[4]), photon counting MKIDs are typical operated in a phase detection mode. The responsivity of an MKID[5] – change in phase per injected quasiparticle - can be approximated as

$$\frac{d\theta}{dN_{qp}} = k(T, f, \Delta, N_0) \frac{\alpha Q_m}{V} \tag{1}$$

where $d\theta/dN_{qp}$ has units of radian per quasiparticle, T is the operating temperature, f is the operating frequency, $\Delta$ is the gap parameter, which is proportional to the critical temperature, $N_0$ is the single spin density of states at the Fermi energy, V is the volume of the superconductor resonator, $Q_m$ is the measured resonator quality factor, and $\alpha$ is the kinetic inductance fraction. *k* relates the change in inductance per quasiparticle to the total inductance and can be calculated from the surface conductivities[6]. When selecting an optimal material, there are a several material-dependent quantities that should be considered. In order to maximize the responsivity, $N_0$, $\Delta$, and V should be minimized while $Q_m$ and $\alpha$ are maximized. For a given material, the most straightforward way to increase the responsivity is to fabricate thinner MKIDs, which will simultaneously decrease the volume and increase the kinetic inductance fraction. However, in some cases the internal quality factor can be reduced in very thin films (e.g., aluminum MKIDs)[6]. The relationship between $\alpha$ and thickness can be seen in equations (2) and (3). The kinetic inductance fraction is the ratio of the inductance that changes with quasiparticle density to the total inductance. It can be calculated[6] as

$$\alpha = \frac{gL_s}{L_m + gL_s} \tag{2}$$

where g is a geometric constant, $L_m$ is the geometric inductance determine by the resonators physical shape, and $L_s$ is the surface inductance. If the film thickness is less than the penetration depth, the surface inductance can be calculated as

$$L_s \approx \left(\frac{\hbar R_S}{\pi \Delta}\right) \tag{3}$$

where we determine Δ, the energy gap, from the critical temperature $T_C$, and the surface resistance $R_S = \rho/t$ from ρ the resistivity and t the film thickness. The thinner the film becomes, the larger $L_s$ and α become.

Although the number of quasiparticles generated by an x-ray photon can be quite large, the decrease in responsivity with increasing film thickness is an issue when designing detectors for photons in the x-ray energy range. In order to create an x-ray detector with high detection efficiency it must have a high x-ray absorption efficiency. The x-ray absorption efficiency is a function of both the detector material density, atomic number, and thickness[7]. To minimize the thickness of the detector, a high density material is required. MKIDs have been made from a wide variety of elemental and alloy superconductors, including Al, Nb, Ta, Mo, Ir, Re, Ti, TiN, NbTi, and NbTiN. Al and TiN are the most commonly used due to high Qs and long lifetimes. Unfortunately, both of these are fairly low density materials (5.4 g/cm³ for TiN or 2.7 g/cm³ for Al) with corresponding attenuation lengths (the thickness at which the absorption efficiency is 1/e) for 6 keV x-rays of 5.5 μm and 33.6 μm respectively. Ta is a dense material (16.7 g/cm³) with a smaller attenuation length (1.8 μm at 6 keV) but has a small kinetic inductance fraction. Given the limited number of alternative materials available, we have searched for unconventional superconducting materials. To this end we have started to examine tungsten silicides, which have densities of 9.8 and 14.5 g/cm³ for $WSi_2$ and $W_5Si_3$ respectively, with attenuation lengths at 6keV of 3.4 and 2.1 μm respectively.

Both amorphous tungsten[8] and tungsten silicide[9] have been used for superconducting electronics and detectors. It has been shown that the $T_C$ of sputtered thin film W can be controlled by deposition condition (e.g. under-layer material or substrate[10]) or doping with silicon. Like TiN, $WSi_x$ has a $T_C$ that depends on stoichiometry and was mapped by Kondo[11] for transition temperatures above 1.9K. For both very small and large Si concentrations the $T_C$ falls below 1.9K (the Tc will decrease to that of pure W – as low as 15mK for the alpha phase – as the silicon content is decreased to zero), but for Si atomic percentages ranging from ~7% up to ~60% the $T_C$ rapidly rises up to 5K with a maximum $T_C$ for Si atomic percentages between 20 and 40%. The normal-state resistivities range from 100-500 μΩcm depending on silicon content. This behavior can be explained in the context of a metal-insulator transition where disorder increases the attractive electron-phonon part of the BCS interaction potential while also decreasing the densities of states.[12,13] These competing factors result in an enhanced $T_C$ from the purely metallic phase. This model can explain the enhanced $T_c$ of several non-granular superconducting systems that exhibit a metal-insulator transition besides $WSi_x$ (e.g., Nb-Ti, $Nb_3Sn$, Mo-Si, Mo-Ge, Mo-Nb, Ta-Ti).

We have deposited thin films of two stoichiometries: $W_5Si_3$ and $WSi_2$. Single crystal $W_5Si_3$ was first reported to have an enhanced $T_C$ of ~2.83K by Hardy and Hulm[14], while Kondo observed a $T_C$ of ~4.5K in amorphous films with 37.5% atomic percent of Si. Although $WSi_2$ is a stable compound (similar to other transition metal silicides, $XSi_2$) it has not been widely studied due to a $T_C$ below 1.9K. Our films were deposited using DC magnetron sputtering onto r-plane sapphire from nominally stoichiometric 99.5% pure powder-hot-pressed targets of $WSi_2$ and $W_5Si_3$. Prior to deposition the wafers were solvent cleaned. The films reported in this letter were deposited at 10 mTorr and 8 mTorr, respectively, for $W_5Si_3$ and $WSi_2$, and XRD was used to confirm the amorphous nature of the films.

We first characterized the critical temperature and resistivity of the films. Resistivity values were determined by room temperature four point probe measurements. $T_c$ measurements were made by measuring the resistance as a function of temperature. Both materials show a sharp normal to superconducting transition. The $W_5Si_3$ film has a $T_C$ of 4K and normal state resistivity of ~ 250 μΩcm and the $WSi_2$ film has a $T_C$ of 1.8K and normal state resistivity of ~ 450 μΩcm. The residual resistance ratios (RRR) are close to 1, due to the amorphous nature of the material. The resistivities are much larger than those of the elemental superconductors and up to 4 times that of TiN, indicating a large surface inductance and thus large kinetic inductance fraction. In order to determine the surface inductance from equation (3), the film thickness must be less than the penetration depth. For a film in the local limit we can calculate the penetration depth from the $T_C$ and normal state resistivity using Mattis-Bardeen theory[15,16]

$$\lambda = \sqrt{\rho_{20K}/T_c} \times 105 \qquad (4)$$

where λ is the penetration depth in nm and $\rho_{20K}$ is the resistivity at 20K in μΩcm. From the values above we calculate penetration depths of 0.83 μm for $W_5Si_3$ and 1.8 μm for $WSi_2$. These large penetration depths mean that $WSi_x$ films should maintain a large kinetic inductance fraction for films up to several microns thick. Additionally, we can use equation (3) to calculate a surface inductance for our films. We calculate a surface inductance of 2.37 pH and 10.5 pH for 350nm films of $W_5Si_3$ and $WSi_2$, respectively. From these values we can estimate the kinetic inductance fraction of ~0.45 and ~0.75 respectively.

Films with thickness ranging from 80nm to 1μm were patterned using optical lithography and reactive ion etching in $SF_6 + O_2$[17] into both quarter wavelength co-planar wavequide (CPW) resonators and lumped element resonators. Unless otherwise noted, the results below are for CPW resonators with a center strip width of 8 μm and a gap width of 7 μm. Resonators were placed in a sample box mounted to the 50mK cold stage of an adiabatic demagnetization cryostat. The sample box is surrounded by a magnetic shield to limit trapped flux during sample cool down and interference from stray magnetic fields. A signal generator excites the resonators and the signal transmitted past the resonator, $S_{21}$, is sent through an IQ mixer to determine the real and imaginary components of $S_{21}$. The resonance frequency is determined by sweeping the frequency of the signal generator and finding a minimum in the magnitude of $S_{21}$. Noise measurements are made by collecting IQ data for 10 seconds at a sampling rate of 250 kHz.

Figure 1 shows the transmission, $S_{21}$, past a $W_5Si_3$ (1a) and a $WSi_2$ (1b) CPW resonator as a function of frequency. Both resonators have internal quality factors of > $10^5$ as determined from a fit of the data using a skewed Lorentzian. The resonators were fabricated using the same process and pattern. A large difference in resonance frequencies is clearly seen (4.7051 GHz for $W_5Si_3$ and 3.0296 GHz for $WSi_2$) when compared to the design (geometric) resonance frequency of ~6.18 GHz. The kinetic inductance fraction can be experimentally calculated according to

$$\alpha = 1 - \left(f_m/f_g\right)^2 \qquad (5)$$

where $f_m$ is the measured resonance frequency and $f_g$ is the geometric resonance frequency[6]. From the measured frequency shifts we calculate $\alpha$ = 0.43 for 350 nm CPW $W_5Si_3$ and $\alpha$ = 0.75 for 350 nm CPW $WSi_2$. The values are consistent with those calculated from the surface inductance and the geometric inductance of our CPW resonators. The large $\alpha$ values are comparable to those of TiN films[3] and much larger than those of Al or Ta (0.01 to 0.05 depending on thickness). We have also tested several lumped element resonators using $WSi_2$, which show $\alpha$ = 0.96 for a 100nm thickness and $\alpha$ = 0.77 for a 1μm thick film.

We measured the frequency noise for CPW resonators made from films of Al, $W_5Si_3$, and $WSi_2$. The frequency noise is a means to compare resonators with different quality factors and resonant frequencies. It can be thought of as the frequency fluctuation necessary to see the measured phase noise, which is the dominant noise in the device. The measured noise depends on the internal power of the resonators and decreases with increasing power. We have measured the frequency noise of each resonator at the same internal power $P_{internal} = 2Q_m^2 P_{readout}/\pi Q_c \sim$ -40dBm. We measure frequency noises at 1kHz of $6.3 \times 10^{-20}$ $Hz^2$/Hz, $2.3 \times 10^{-20}$ $Hz^2$/Hz, and $5.7 \times 10^{-20}$ $Hz^2$/Hz for 350nm films of Al, $W_5Si_3$, and $WSi_2$ respectively. These levels are comparable to frequency noises measurements of comparable CPW resonator geometries of Al[18] and TiN[19]

The large quality factors and kinetic inductance fractions mean that $WSi_x$ should have a large response. In addition, the $WSi_x$ alloy system is predicted to have a small single spin density of states ($N_0 = 5.83 \times 10^9$ μm$^{-3}$ eV$^{-1}$ for $WSi_2$ and $N_0 = 2.17 \times 10^9$ μm$^{-3}$ eV$^{-1}$ for $W_5Si_3$)[12] compared to TiN ($N_0 = 8.7 \times 10^9$ μm$^{-3}$ eV$^{-1}$)[19], Al ($1.72 \times 10^{10}$ μm$^{-3}$ eV$^{-1}$)[12], or Ta ($N_0 = 7.38 \times 10^{10}$ μm$^{-3}$ eV$^{-1}$)[12]. From these values we can use equation (1) to compare the theoretical response from resonators of the same volume composed of each material. In Figure 2 we show the phase response from resonators with a volume of $5 \times 10^4$ μm$^3$ and a $Q_m = 1 \times 10^4$. The $WSi_2$ offers the largest response followed by $W_5Si_3$ and TiN. We have fabricated lumped element resonators from 1 μm thick $WSi_2$ and 300nm thick Ta and exposed them to X-rays from Fe$^{55}$ (~ 6keV) and Cd$^{109}$ (~22keV) sources. We chose to compare $WSi_2$ and Ta given their comparable x-ray absorption efficiency. A pulse using Fe$^{55}$ from the $WSi_2$ resonator is shown in Figure 3. The phase pulse height of ~ 150 degrees is less than calculated, likely due to saturating the resonator with this volume. The 300nm Ta resonator has pulse heights of ~ 30 degrees from the Cd$^{109}$ source as expected from equation (1). However, the x-ray absorption efficiency for 300nm of Ta is ~ 15% compared to ~25% for 1μm of WSi2; increasing the thickness of Ta for better absorption efficiency will only decrease the response.

Finally, we have investigated the quasiparticle lifetime ($\tau_{qp}$) of the $WSi_x$ films by illuminating the MKIDs with x-rays. The accurate measurement of the non-equilibrium quasiparticle population (i.e., the energy resolution) is proportional to $1/\sqrt{\tau_{qp}}$[20]. We observed decay times for both materials of ~ 5μs. We would have expected longer lifetimes in the $WSi_2$ resonators due to the lower $T_c$[21]. These decay times are on the order of the resonator decay time, $\tau_{res} = 2Q_m/f_r$, and with the lack of a variation with $T_c$ suggests that the lifetime is shorter than the ~ 5μs resonator decay time. These lifetimes are much shorter than those of other common materials (100 μs for 1.1K TiN and > 1ms for Al) and indicate the quasiparticle lifetime may be dominated by self-recombination or inelastic losses due to defects[22]. The short lifetimes will require resonators with a small $Q_m$ (~ $10^3$) to avoid roll off of the phase response due to the decay time. Recent work[23] has shown that lifetimes in Al can be increase by placing the resonator on a SiN membrane to control the loss of athermal phonons from quasiparticle recombination. We plan to investigate how the quasiparticle lifetime of $WSi_x$ is affected by substrate choice and how this will affect the energy resolution.

In summary we have examined two compositions of the tungsten-silicon alloy – $W_5Si_3$ and $WSi_2$ - for use as microwave kinetic inductance detectors. We examined the critical temperature and resistivity of thin films, and the quality factors and noise properties of quarter wavelength CPW resonators. The alloy system has many positive features including a $T_C$ that varies with composition from 0 to 4K, a high normal state resistivity up to ~ 450 μΩcm and a high kinetic inductance faction (0.77 for a 1μm thick $WSi_2$ film). We have fabricated both CPW and lumped element resonators with quality factors >$10^5$ (often > $10^6$). The short lifetimes may present a barrier for use as a MKID, but this might be overcome by placing the devices on a SiN membrane. For use as an x-ray detector, the high density greatly reduces the required thickness of the material without significantly degrading the kinetic inductance fraction. X-ray induced phase pulses in lumped element resonators of $WSi_2$ show a large response. This material system offers promise for optical detectors due to its high resistivity and kinetic inductance fraction and for x-ray detectors due to its high x-ray absorption efficiency and phase response that may allow the realization of a detector with film thickness of a micron or greater.

The authors would like to thank Lisa Gades, Ralu Divan, Suzanne Miller, Brandon Fisher, and Yejun Feng. Use of the Center for Nanoscale Materials was supported by the U. S. Department of Energy, Office of Science, Office of Basic Energy Sciences, under Contract No. DE-AC02-06CH11357. Work at Argonne National Laboratory was supported by the U. S. Department of Energy, Office of Science, Office of Basic Energy Sciences, under Contract No. DE-AC02-06CH11357.

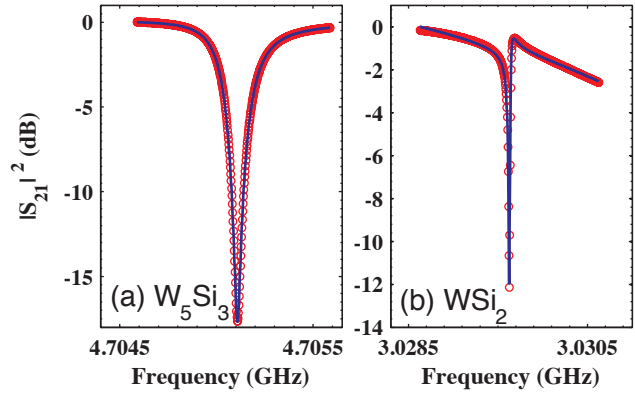

Figure 1: (Color online) Transmission, $S_{21}$, past a quarter wavelength CPW resonator of a) 350nm $W_5Si_3$ at $P_{int}$ = -44.1 dBm with $F_r$ = 4.7051 GHz, $Q_m$ = 2.30 × $10^4$ and $Q_i$ = 1.76 × $10^5$ and b) 350nm $WSi_2$ at $P_{int}$ = -39.3 dBm with $F_r$ = 3.0296 GHz, $Q_m$ = 8.38 × $10^4$ and $Q_i$ = 3.32 × $10^5$. Measured data are shown with open circles and the line is a fit to the data using a skewed Lorentzian. Note the shift to lower frequencies due to large kinetic inductance fraction when compared to the design resonant frequencies of ~6.18 GHz.

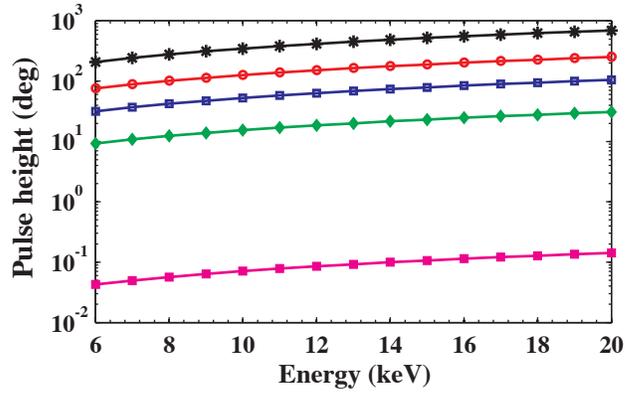

Figure 2: (Color online) Calculated phase shift as a function of photon energy for 1μm thick resonators with $Q_m = 10^4$, Volume = $5 \times 10^4$ μm$^3$, and $f_r$ = 6 GHz using WSi$_2$ ($\alpha$ = 0.75, *), W$_5$Si$_3$ ($\alpha$ = 0.43, O), TiN ($\alpha$ = 0.75, □), Al ($\alpha$ = 0.05, ◆), and Ta ($\alpha$ = 0.01, ■).

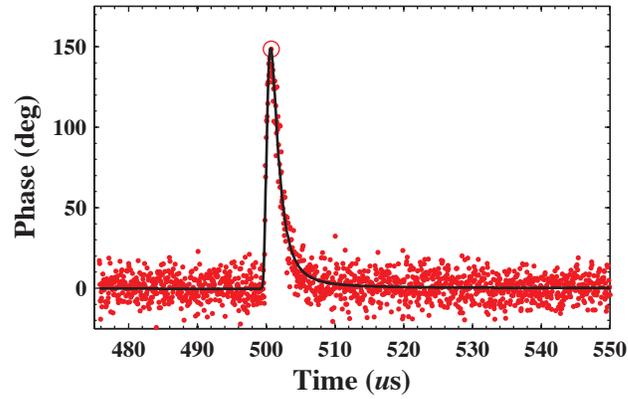

Figure 3: (Color online) Phase pulse from an $Fe^{55}$ x-ray absorption in a 1 μm thick $WSi_2$ resonator with $Q_m = 2 \times 10^4$ and $\alpha = 0.75$. The pulse is smaller than calculated due to saturation of the detector at high phase deviations. The black line is semi-Gaussian fit to the data.